\documentclass[aps,prl,superscriptaddress,amsmath,amssymb,noshowkeys,twocolumn,showpacs]{revtex4-1}
\usepackage{graphicx}
\usepackage{dcolumn}
\usepackage{bm}
\usepackage[usenames]{color}
\usepackage{epstopdf}

\newcommand{\affchic}{\affiliation{James Franck Institute and Department of Physics, The University of Chicago, Chicago, Illinois 60637, USA}}
\newcommand{\affpof}{\affiliation{Physics of Fluids Group, University of Twente, P. O. Box 217, 7500 AE Enschede, The Netherlands}}

\begin{document}

\title{Dense Suspension Splat:  Monolayer Spreading and Hole Formation After  Impact}

\author{Luuk~A.~Lubbers}
\affchic
\affpof
\author{Qin~Xu}
\affchic
\author{Sam~Wilken}
\affchic
\author{Wendy~W.~Zhang}
\affchic
\author{Heinrich~M.~Jaeger}
\affchic

\begin{abstract}
We use experiments and minimal numerical models to investigate the rapidly expanding monolayer formed by the impact of a dense suspension drop against a smooth solid surface. The expansion creates a lace-like pattern of particle clusters separated by particle-free regions. Both the expansion and the development of the spatial inhomogeneity are dominated by particle inertia, therefore robust and insensitive to details of the surface wetting, capillarity and viscous drag.
\end{abstract}



\maketitle

Since the pioneering work by Worthington~\cite{worthington1876forms} the spreading of liquids droplets upon impact has remained an active research area~\cite{rein1993phenomena,marengo2011drop}. At meters-per-second impact speeds, the spreading divides into two stages~\cite{clanet2004maximal}:  an initial, rapid spreading dominated by liquid inertia and, consequently, insensitive to surface wetting, capillary or liquid viscosity, followed by a slower evolution where the intricate interplay of these effects is important.  

Here we examine an analogous inertia-dominated spreading dynamics in dense suspension impact.  We use a suspension of rigid, non-Brownian particles at high volume fraction ($60 \%$ or above). This impact regime has received little study~\cite{peters2013does}. Previous studies on particle-laden drop impact have mainly analyzed slow evolution in dilute and semi-dilute suspensions~\cite{de2007marangoni,hu2006marangoni,nikolov2002superspreading,nicolas2005spreading}.  We find that impact at several meters per second produces a novel outcome~(Fig.~1):  the suspension drop deforms into a splat comprised of a single layer of densely packed particles immersed in a thin liquid layer. As the splat expands, void-like regions appear and grow, causing the final splat to display a lace-like pattern of particle clusters separated by particle-free regions.  Because particle inertia dominates the expansion and the instability, the monolayer splat dynamics is robust and only weakly modified by surface wetting, capillary and viscous drag.
\begin{figure}[h!]
	\centering
	\includegraphics[width=0.4\textwidth]{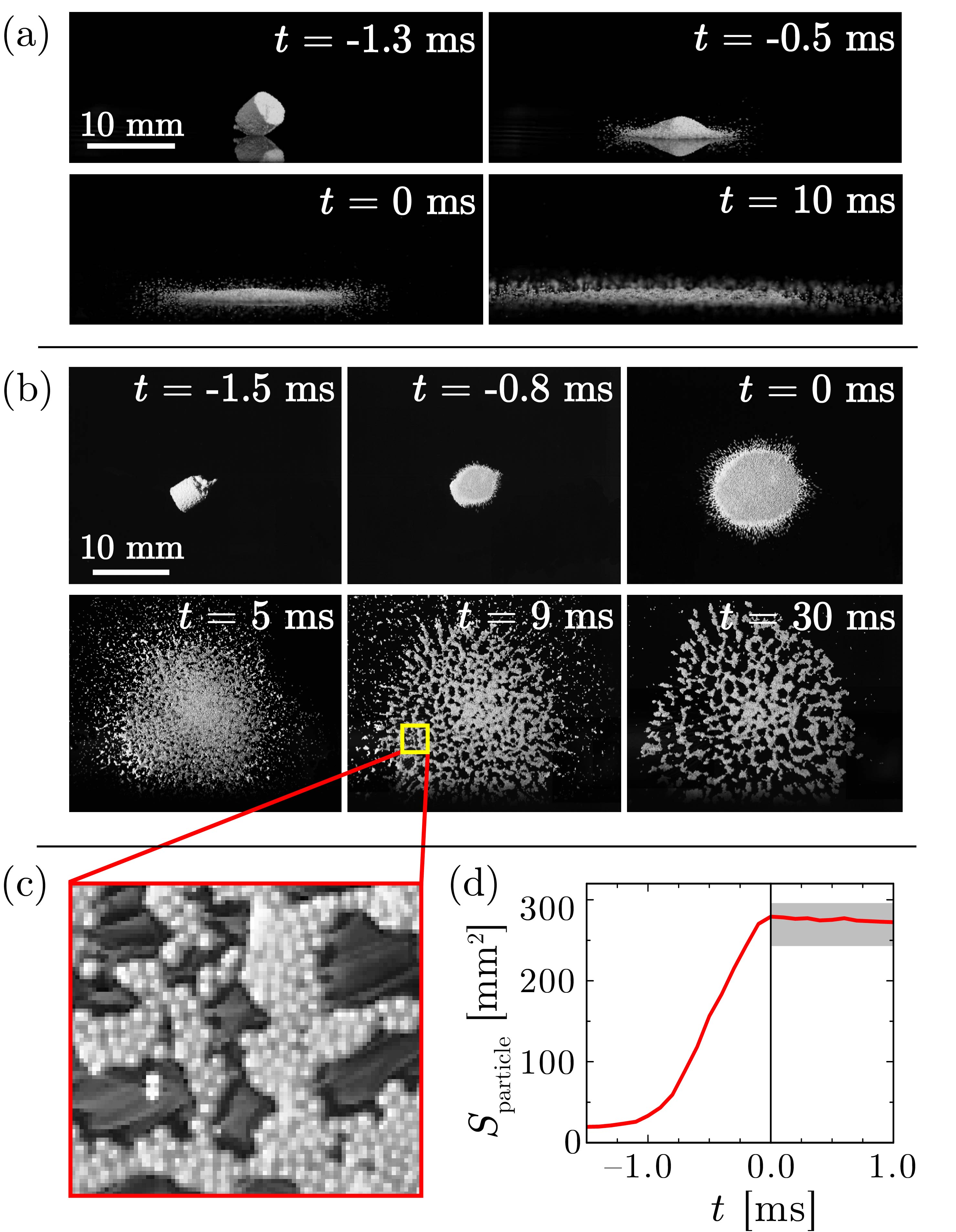}
	\caption{
(Color online) Dense suspension impact, splat, and instability. (a)~Side view: A cylindrical plug impacts a smooth dry glass surface, splashes by ejecting particles upwards and flattens into a monolayer. (b)~Bottom view: The initial, nearly circular splat expands.  Inhomogeneities appear as regions of particle clusters separated by particle-free regions (dark). (c)~Close-up: clusters drag streaks of liquid along as they move outwards, visible by the contrast in liquid color~\cite{movie}. (d)~Substrate area coverage as function of time. Once the covered area approaches a constant value (shaded region),  particles are spread out in a monolayer.
}
\label{fig:fig1}
\end{figure}

This insensitivity to material and surface properties is often the desired outcome in coating processes. This makes our results useful in ongoing efforts to assess the cohesive strengths of colloidal semiconductor quantum dots by measuring their maximal splat size after impact~\cite{qi2011impact}, as well as applications such as thermal spray coating of sintered powders~\cite{fukumoto1999flattening} and additive manufacturing using inkjet printing~\cite{seerden2004ink,derby2011inkjet}. These processes often involve concentrated suspensions. Moreover the impact speeds are often very large. As a result, despite the smaller particles used in these processes, the post-impact spreading dynamics belongs in the same particle-inertia dominated regime as our experiments.
  
\paragraph{Experiments ---} 
Dense suspensions were made by adding spherical $\mathrm{ZrO_{2}}$ particles  $\left( \mathrm{d} = 250 \pm 22\mathrm{\; \mu m,\;}\mathrm{\rho_p} = 5.68 \times 10^{3} \mathrm{\;kg \cdot m^{-3}}\right)$ to water or silicon oils. Letting the particles sediment inside a straight cylindrical syringe produces packing fractions of $\phi=0.61\pm 0.02$. As gravity pulls the suspension down, a pinch-off occurs below the cylinder opening~\cite{miskin2012droplet,bertrand2012dynamics}. In the dense suspension limit studied here, the plugs preserve the cylindrical shape of the syringe, resulting in a plug radius $\left( R_p \approx 2.25\pm0.05\mathrm{\;mm}\right)$, and have a height ~$L\sim2R_p$. The substrate was a smooth, horizontal glass plate $1.6\pm 0.03 \mathrm{\;m}$ below the syringe. 

Figure~\ref{fig:fig1} shows typical image sequences of the impact, recorded by high-speed video. We denote $t=0$ as the moment when a monolayer first forms.  Before this moment the cylinder-shaped plug flattens into a single-particle layer.  This time point can be defined precisely by viewing plug impact onto a transparent glass slide from below and plotting the substrate coverage area as a function of time [Fig.~\ref{fig:fig1}(c)].  The transition to  constant area indicates the monolayer onset.

\begin{figure}[t!]
\centering
\includegraphics[width=0.39\textwidth]{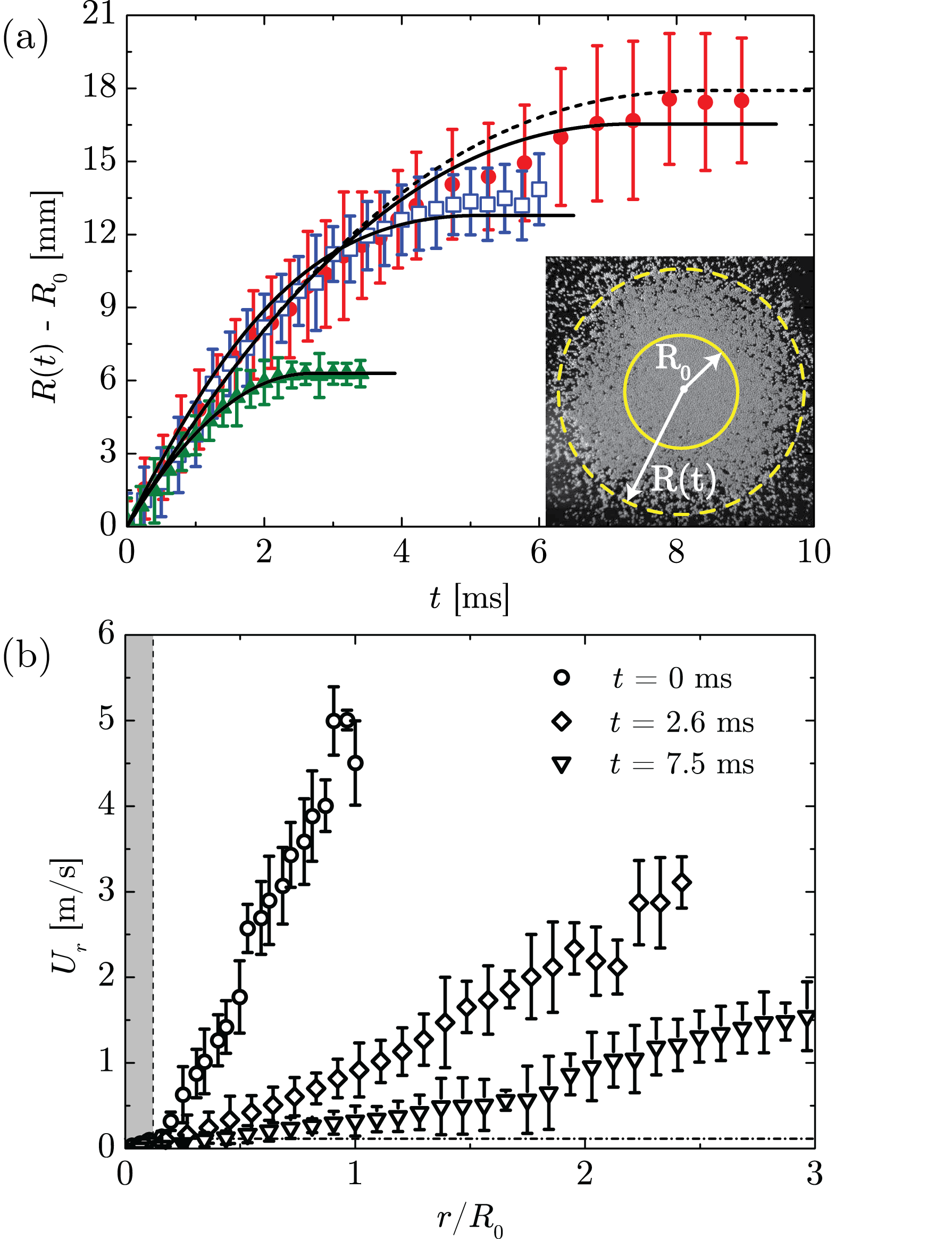}
\caption{
(Color online) Splat expansion dynamics. (a)~Expansion radius, defined as the difference between the splat's outer edge $R(t)$ and the initial monolayer radius $R_0$. All data for $\mathrm{ZrO_{2}}$ particles in water ($\blacktriangle$), and two silicone oils with lower surface tension and viscosity values of 1.8 cSt~($\bullet$) and 10 cSt~($\square$). Predictions from the leading-edge model (solid lines) and the chain model (dashed line) are shown. (b)~Radial velocity profiles of the $\mathrm{ZrO_{2}}$-in-silicon-oil splat at the moment of formation $t=0$ and 2 instances afterwards. The dot-dashed line marks the velocity $U_r^* $ where $\rho_p \left( U_r^* \right)^2 d/\gamma = 1$. 
}
\label{fig:fig2}
\end{figure}

After $t=0$, the monolayer expands radially and develops holes. We measure the expansion by azimuthally averaging the particle density as a function of radial distance and define the splat's edge $R(t)$ as the sharp transition zone from high to zero density. Fig.~\ref{fig:fig2}(a) plots $R(t) - R_0$, the difference between the splat radius and the initial monolayer radius $R_0$.  The velocity field $U_{r}\left(r\right)$ is obtained by azimuthally averaging the particle motion [Fig.~\ref{fig:fig2}(b)]. We find a linear straining flow, starting at zero velocity at the small dead zone of immobile particles at the center of impact (shaded region). At later times, this linear straining flow weakens but retains its form. This velocity field is consistent with the self-similar, inertia-dominated thin-film spreading flow after impact, first described by Yarin \& Weiss~\cite{yarin1995impact}, and supports the idea that inertia dominates the expansion. We quantify the time evolution of the spatial inhomogeneity in terms of the area fraction in the splat occupied by the particle-free regions. Since the instability grows fastest near the outer edge and slower in the interior, we divide the splat into an inner and outer annulus that contain approximately the same particles over time, and plot the average area fraction of void regions within each annulus as a function of time. The measured void fraction initially grows rapidly, then slows and saturates~[Fig.~\ref{fig:fig4}(b)].

\begin{figure}[t]
\includegraphics[width=0.45\textwidth]{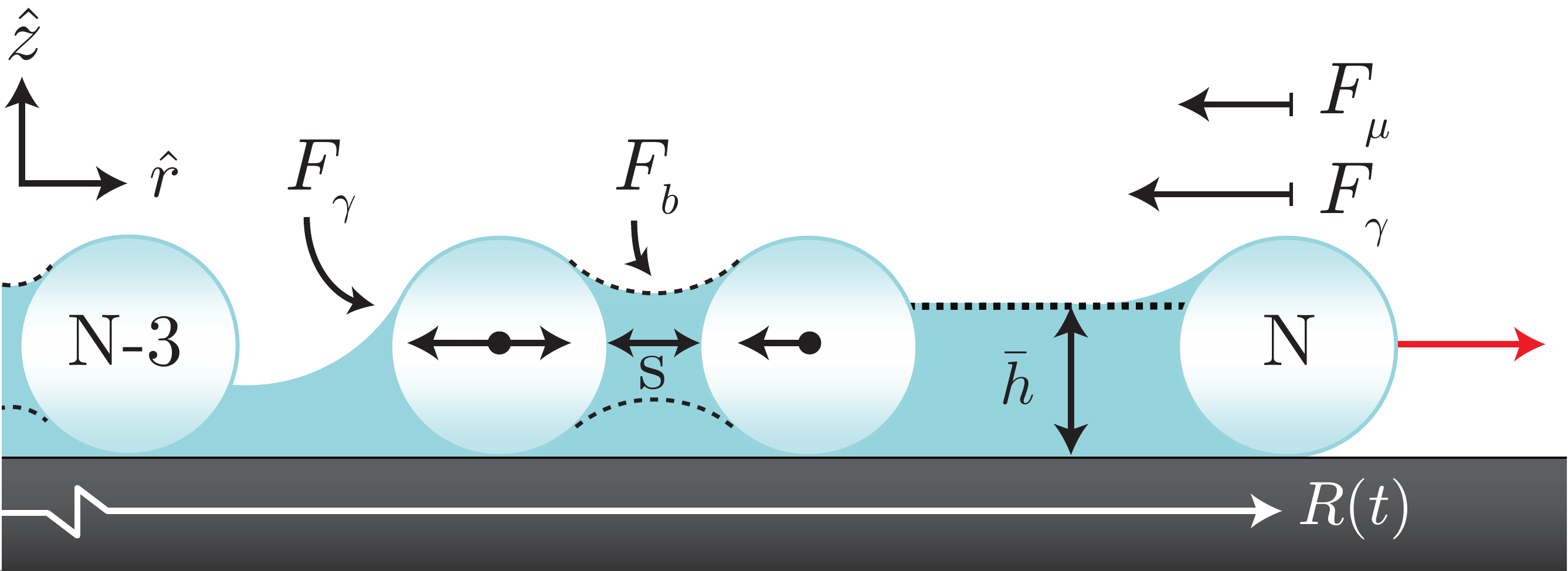}
\caption{
(Color online) In the leading-edge model for splat expansion, the splat edge, initially ejected with horizontal speed $\dot{R}(t=0)$, slows over time due to resistance by surface tension $F_{\gamma}$ and drag force $F_\mu$ from a trailing liquid film. In the chain model for splat instability particle-free regions emerge from variations in the initial radial velocity field. Beyond a critical separation $s_c$ between adjacent particles, bridge-like menisci transform into trailing liquid streaks and the force switches from a bridging force $F_b$ to a trailing streak resistance $F_\gamma$ acting solely on the faster moving particle in front. 
}
\label{fig:fig3}
\end{figure}
Our experiments are characterized by the particle-based Weber number $We_p \equiv \rho_p d U^2_0/ \gamma$, where $\gamma$ is surface tension and $U_0 = \dot{R}(t=0)$ is the initial expansion speed for the monolayer, and the Stokes number $St \equiv \rho_p d U_0/ \mu$, where $\mu$ is the suspending liquid viscosity. Using speeds at the expanding edge, the water and silicone oil suspension impacts featured in ~\ref{fig:fig2}(a) have $We_p \approx 520$ (water) and $1900$ (silicone oil), and $St$ values of $7400$ (water), $4000$ (1.8 cSt oil) and $800$ (10 cSt oil). The rate of strain is so large that $We_p$ and $St$ are both much larger than $1$ over almost all the splat interior. 

\paragraph{Splat Expansion---}
Given $We_p \gg 1$ and $St \gg 1$ we consider the following particle-inertia dominated model.  We assume the splat expands as fast as particles at the splat's leading edge can move and that these particle motions are unaffected by collisions.  As a result, the only forces acting on the leading-edge particles are surface tension and drag due to motion relative to the liquid layer and/or the solid substrate.  First, to estimate the force due to surface tension, we note that in the experiments the particle remains fully coated by the suspending liquid through out the expansion. This gives rise to an asymmetric free surface shape: a thin liquid film coats the front surface of the particle at the outermost edge while the rear half of the particle remains immersed in a thicker liquid layer (Fig.~\ref{fig:fig3}). This yields a retarding force $F_\gamma = \alpha \pi \gamma d /2$. Because the value of the constant $\alpha$ depends on the free-surface asymmetry, it varies from particle to particle and changes over time depending on the precise configuration of neighbors~\cite{wettingComment, supplementalMaterials}. This makes an explicit calculation cumbersome. Here we will simply determine the value of $\alpha$ by fitting the model predictions against measurements. Second, the drag experienced by a leading-edge particle moving outwards with speed $\dot{R}(t)$ has several distinct contributions. Measurements suggest the dominant contribution $F_\mu$ is viscous drag due to a thin trailing liquid streak~\cite{supplementalMaterials}. If the average liquid layer thickness in the splat is $\bar{h}$, then the average viscous stress experienced by the particle as it drags a liquid streak along is $\mu \dot{R}/\bar{h}$. If we assume in addition that this viscous stress acts over $\pi d^2/4$, the projection of the particle surface area in the direction of motion, then $ F_{\mu}$, the total drag due to the trailing streak, is  $( \mu \dot{R} / \bar{h}) (\pi d^2 / 4)$. Comparison with measured data presented later will show that this expression gives a quantitatively correct description for splat expansion at high liquid viscosities.  

Requiring $ma = F = -F_{\gamma} - F_\mu$ where $m$ is the particle mass and $a$ its acceleration at the leading edge yields an evolution equation for the splat radius $R(t)$
\begin{equation}
\rho_p \left( \frac{\pi d^3}{6} \right) \ddot{R}= - \left( \mu \dot{R}/\bar{h} \right) \left(\frac{\pi d^2}{4} \right) -{\alpha\gamma \frac{\pi d}{2}} \ . 
\label{eq:rmom}
\end{equation}
Since the volume of liquid inside the suspension is conserved over time and the liquid layer is much thinner than the particle diameter, the unknown liquid layer thickness $\bar{h}(t)$ is directly related to $R(t)$ via $(1 - \phi) V_p = \pi R^2(t) \bar{h}(t)$, where $V_p = \pi R^2_p L$ is the volume of the suspension plug. 

Fig.~\ref{fig:fig2}(a) plots the measured splat expansion dynamics against those calculated using Eqn.~(\ref{eq:rmom}) initialized with measured values of $R(t=0)$ and $\dot{R}(t=0)$.  We found that choosing the prefactor $\alpha$ to be $2.9$ yields the best agreement with the measured evolution for water-solvent suspension,  which has the highest surface tension value~\footnote{Varying $\alpha$ by 10\% still allows a fit within the error bar range of experiments.}.  The model also produces good agreement with data from the $10$ cst silicone oil suspension, where the expansion is slowed by viscous drag.  This shows that the proposed expression for $F_\mu$ is quantitatively accurate. As far as we are aware, this simple drag law has neither been proposed nor tested against data in previous studies. 

\paragraph{Splat Instability---}
Because the monolayer splat regime is characterized by large particle inertia together with small surface tension and viscous drag, the observed spatial inhomogeneity originates as small variations in the particle velocities within the initial, densely packed monolayer splat. These imperfections are amplified by the subsequent rapid expansion and grow into a lace pattern.  This instability is qualitatively different from capillarity induced aggregation~\cite{vella2005cheerios}, which proceeds on a time-scale far longer than the monolayer expansion time-scale.  The inertia-dominated instability is also far less sensitive to the detailed forms of capillarity and/or viscous drag.  Neither is required to nucleate the instability. Nor do they control the instability growth rate. As a result, a minimal numerical model in which particle inertia is weakly perturbed by capillary and viscous drag is capable of quantitatively reproducing the main features of the instability.

A one-dimensional (1D) model based on this scenario gives reasonable agreement with measured growth rates for the spatial inhomogeneity. The model considers a chain of $N$ particles which lie along a ray emanating from the center of the splat (Fig.~\ref{fig:fig3}). Because surface tension and viscous drag merely perturb the dominant inertial motion, simple approximations will turn out to be sufficient for a quantitatively accurate description. Specifically, each particle in the chain experiences viscous drag $F_\mu = (\mu \dot{R}_i / \bar{h}) (\pi d^2/4)$ where $\dot{R}_i$ is the speed of the $i$th particle in the chain. Initially the particles in the chain are closely packed together and each experiences cohesive capillary forces $F_b$ with neighbors ahead of and behind itself. As the splat expands rapidly, the interface profile is dominated by particle inertia therefore deforming to coat the particles as they move outwards. In the region between particles, the highly dynamical surface shape is controlled by liquid inertia and viscous drag, and therefore remains nearly flat once the particles are sufficiently far apart. Therefore no cohesive capillary forces are expected to be present between particles more than a critical distance $s_c$ apart. To model this force, we use the formula for the cohesion exerted by an axisymmetric, static liquid bridge connecting two fully wetted spheres~\cite{herminghaus2005dynamics}. This is not because this corresponds to our dynamic situation, but because it recapitulates the main desired features once $s_c$ is allowed to vary~\cite{supplementalMaterials}. As the gap between the neighboring particles exceed a critical value $s_c$, the cohesive capillary interaction switches off ($F_b =0$). Instead, motivated by images from the experiment showing faster moving particles leaving streaks of liquid behind themselves, we require that a particle far ahead of its neighbor in the chain model experiences a retarding force due to surface tension $F_{\gamma} = \alpha \pi \gamma d/2$, while the left-behind neighbor no longer is pulled forward force by a liquid bridge.

The dashed line in Fig.~\ref{fig:fig2}(a) gives the position of the outermost particle in the $N$-particle chain and agrees well with the measured evolution.  Comparisons with the other two systems also show good agreement and are given in~\cite{supplementalMaterials}.  Finally, we calculate the void fraction evolution from the chain model and plot the results in~\ref{fig:fig4}(b). Importantly, the calculated instability dynamics is robust when changing model parameters. Altering the value for $s_c$ by  $40\%$ from $d/4$ used in generating the chain model result presented in Fig.~\ref{fig:fig4}(b), or using an initial velocity fluctuation that is half, or double the $10\%$ value used, produces negligible changes.

\begin{figure}[t!]
	\centering
	\includegraphics[width=0.39\textwidth]{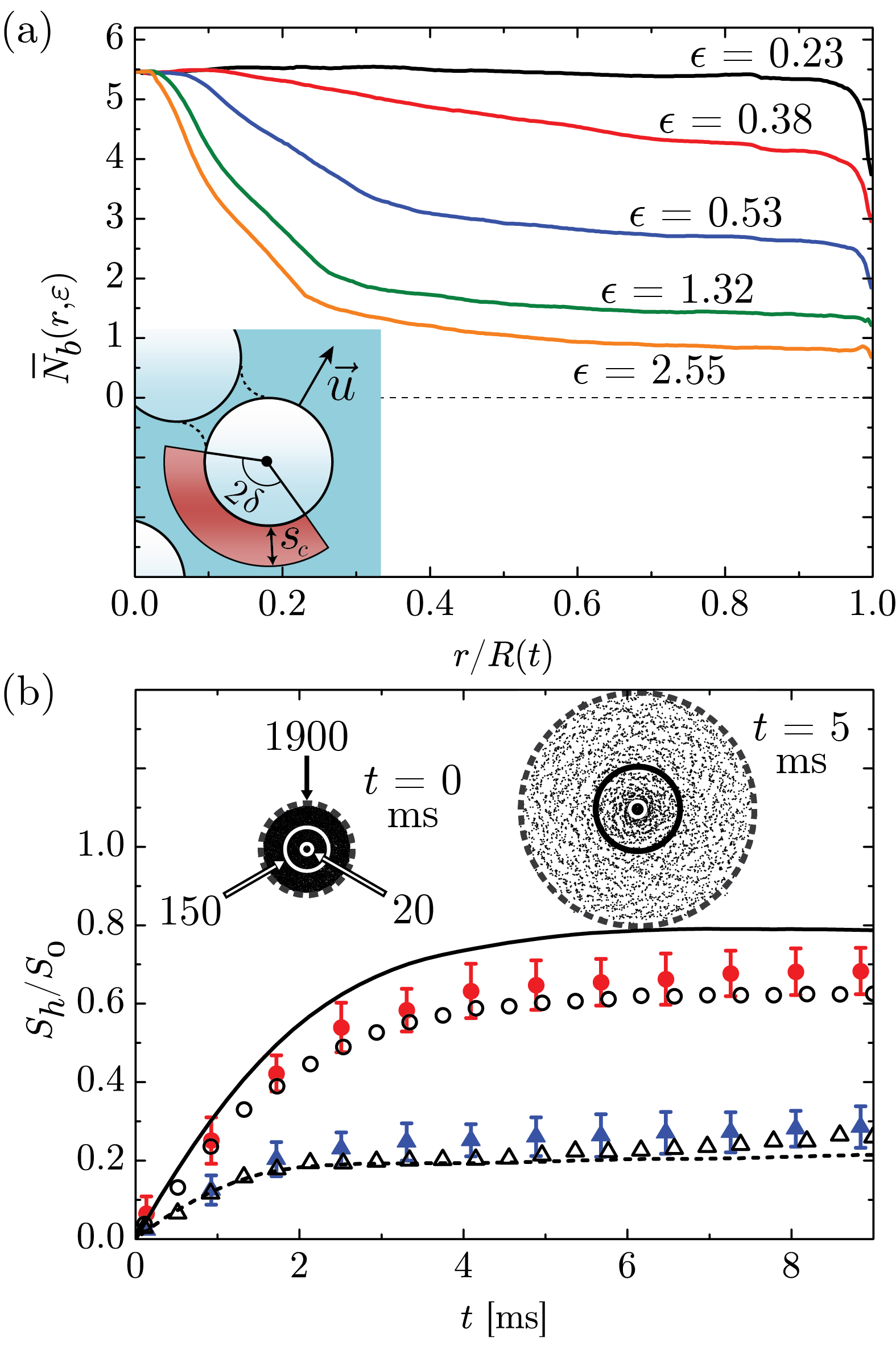}
	\caption{
(Color online) Instability dynamics. (a)~Average number of capillary-bridge bonds per particle, $\bar{N}_b$, as function of $r/R(t)$, the radial distance normalized by current splat radius, for different values of radial strain $\epsilon = [R(t) - (R_0 - R_{\rm DZ})]/\left( R_0 - R_{\rm DZ}\right)$. Inset: 2-dimensional generalization of $F_b$ and $F_\gamma$. A cohesive capillary bridge bond between neighboring particles becomes a trailing streak if the neighboring particle lies outside a wedge of opening angle $2\delta$ and radius $\left( d/2 \right)+s_{\rm c}$ (shaded region). (b)~Area fraction of particle-free regions in circular annuli within the splat as a function of time. The boundary between the inner and outer annuli is chosen to lie at $\rho U_{b}^2 d/\gamma = 150$, where $U_{b}$ is the initial speed of the particle at the boundary. Experiments ($\bullet$ \ $\blacktriangle$), one-dimensional chain model (solid and dashed lines), and two-dimensional numerical model ($\circ$ \ $\triangle$) agree quantitatively. Inset: snapshots from the simulation.  
}
	\label{fig:fig4}
\end{figure}

\paragraph{2D Simulation of Instability---}
We next refine the chain model by going to 2D by prescribing the same $F_\mu$ but generalizing $F_b$ and $F_\gamma$ (inset to Fig.~\ref{fig:fig4}(a) and~\cite{supplementalMaterials}) to include capillary interactions with all nearest neighbors, not only those along a radial direction. The inset in Fig.~\ref{fig:fig4}(b) shows two snapshots from the simulation: initially the splat is so densely packed that it appears uniformly black. As the expansion proceeds, voids appear and grow, with the growth rate being faster in the outer regions.  In Fig.~\ref{fig:fig4}(b) we also plot the void area fraction calculated from the simulation.  Including the interaction with azimuthal neighbors allows the 2D simulation to track the initial void growth rate more accurately than the chain model.  This results in a noticeably better fit to the measured evolution.

The simulation also allows us to test our starting assumption that the radial expansion causes the particle dynamics in the monolayer splat to be decoupled, thereby rendering the splat evolution simple. In Fig.~\ref{fig:fig4}(a) we plot $\bar{N}_b$, the average number of nearest neighbors experiencing cohesive capillary interaction, as a function of normalized radial distance.  The different curves correspond to different radial strain $\epsilon = [R(t) - (R_0 - R_{\rm DZ})]/\left( R_0 - R_{\rm DZ}\right)$, with $R_{DZ}$ the radius of the dead zone. This quantity $\bar{N}_b$ is difficult to extract from the experiment but gives direct insight into the degree of collective interactions present. Initially particles everywhere in the splat are densely packed and have on average $5.5$ neighbors. As the radial expansion proceeds and the radial strain grows large, many particles lose cohesive capillary interactions with nearest neighbors, particularly those along the azimuthal direction. This effect is most pronounced near the outer edge, where the expansion speed is the largest. Finally, as the monolayer splat expansion slows, a large outer area in the splat are occupied by particles experiencing one cohesive capillary bond on average. 

\paragraph{Conclusions---} We report a novel outcome of dense suspension impact onto a smooth solid:  the formation of a monolayer splat. Using experiments and minimal numerical models, we analyze the splat expansion and instability dynamics. The quantitative agreement between measurements and model results demonstrate that particle inertia dominates both processes. In this $We_p \gg 1$ and $St \gg 1$ regime, the detailed forms of surface wetting, capillarity and viscous drag have little effect on either the expansion or the instability. This is also a regime relevant for many technologically relevant applications~\cite{fukumoto1999flattening,seerden2004ink,derby2011inkjet,qi2011impact}. The high impact speeds used in these processes ensure that particle inertia remains important despite the smaller particles used. 
Understanding the mechanism responsible for monolayer formation raise more complex questions such as how impact destroys 3D particles clusters, or whether a qualitatively different dynamic appears at lower volume fractions due to long-range viscous flow coupling~\cite{ward2009experimental,timberlake2005particle,colosqui2013hydrodynamically,guazzelli2011fluctuations}. The monolayer spreading dynamics elucidated here provides a particularly simple, yet solid, starting point for tackling these issues. This is because the inertia-dominated expansion rapidly switches off capillary and viscous interactions between neighboring particles, thus making it possible to model and predict the spreading dynamics without having to resolve the considerable complication of suspension rheology~\cite{stickelPowellreview05,mwyartpnas12,catesPRL09, clementMesoscale,wagnerRheologyreview, pouliquenPRL, capillarySuspensionSci12,ebrownJRheo12}.  

\paragraph{Acknowledgements---}
\acknowledgements{We thank I.R.~Peters and M.Z.~Miskin for discussions. We also thank D.~Lohse and J.H.~Snoeijer for bringing Chicago and Enschede together. This work was supported by NSF through its MRSEC program (DMR-0820054) and fluid dynamics program (CBET-1336489).} 


\begin{thebibliography}{34}
\makeatletter
\providecommand \@ifxundefined [1]{%
 \@ifx{#1\undefined}
}%
\providecommand \@ifnum [1]{%
 \ifnum #1\expandafter \@firstoftwo
 \else \expandafter \@secondoftwo
 \fi
}%
\providecommand \@ifx [1]{%
 \ifx #1\expandafter \@firstoftwo
 \else \expandafter \@secondoftwo
 \fi
}%
\providecommand \natexlab [1]{#1}%
\providecommand \enquote  [1]{``#1''}%
\providecommand \bibnamefont  [1]{#1}%
\providecommand \bibfnamefont [1]{#1}%
\providecommand \citenamefont [1]{#1}%
\providecommand \href@noop [0]{\@secondoftwo}%
\providecommand \href [0]{\begingroup \@sanitize@url \@href}%
\providecommand \@href[1]{\@@startlink{#1}\@@href}%
\providecommand \@@href[1]{\endgroup#1\@@endlink}%
\providecommand \@sanitize@url [0]{\catcode `\\12\catcode `\$12\catcode
  `\&12\catcode `\#12\catcode `\^12\catcode `\_12\catcode `\%12\relax}%
\providecommand \@@startlink[1]{}%
\providecommand \@@endlink[0]{}%
\providecommand \url  [0]{\begingroup\@sanitize@url \@url }%
\providecommand \@url [1]{\endgroup\@href {#1}{\urlprefix }}%
\providecommand \urlprefix  [0]{URL }%
\providecommand \Eprint [0]{\href }%
\providecommand \doibase [0]{http://dx.doi.org/}%
\providecommand \selectlanguage [0]{\@gobble}%
\providecommand \bibinfo  [0]{\@secondoftwo}%
\providecommand \bibfield  [0]{\@secondoftwo}%
\providecommand \translation [1]{[#1]}%
\providecommand \BibitemOpen [0]{}%
\providecommand \bibitemStop [0]{}%
\providecommand \bibitemNoStop [0]{.\EOS\space}%
\providecommand \EOS [0]{\spacefactor3000\relax}%
\providecommand \BibitemShut  [1]{\csname bibitem#1\endcsname}%
\let\auto@bib@innerbib\@empty
\bibitem [{\citenamefont {Worthington}(1876)}]{worthington1876forms}%
  \BibitemOpen
  \bibfield  {author} {\bibinfo {author} {\bibfnamefont {A.}~\bibnamefont
  {Worthington}},\ }\href@noop {} {\bibfield  {journal} {\bibinfo  {journal}
  {Proc.~R.~Soc.~A}\ }\textbf {\bibinfo {volume} {25}},\ \bibinfo {pages} {261}
  (\bibinfo {year} {1876})}\BibitemShut {NoStop}%
\bibitem [{\citenamefont {Rein}(1993)}]{rein1993phenomena}%
  \BibitemOpen
  \bibfield  {author} {\bibinfo {author} {\bibfnamefont {M.}~\bibnamefont
  {Rein}},\ }\href@noop {} {\bibfield  {journal} {\bibinfo  {journal}
  {Fluid~Dyn.~Res.}\ }\textbf {\bibinfo {volume} {12}},\ \bibinfo {pages} {61}
  (\bibinfo {year} {1993})}\BibitemShut {NoStop}%
\bibitem [{\citenamefont {Marengo}\ \emph {et~al.}(2011)\citenamefont
  {Marengo}, \citenamefont {Antonini}, \citenamefont {Roisman},\ and\
  \citenamefont {Tropea}}]{marengo2011drop}%
  \BibitemOpen
  \bibfield  {author} {\bibinfo {author} {\bibfnamefont {M.}~\bibnamefont
  {Marengo}}, \bibinfo {author} {\bibfnamefont {C.}~\bibnamefont {Antonini}},
  \bibinfo {author} {\bibfnamefont {I.~V.}\ \bibnamefont {Roisman}}, \ and\
  \bibinfo {author} {\bibfnamefont {C.}~\bibnamefont {Tropea}},\ }\href@noop {}
  {\bibfield  {journal} {\bibinfo  {journal}
  {Curr.~Opinion~in~Colloid~Inter.~Sci.}\ }\textbf {\bibinfo {volume} {16}},\
  \bibinfo {pages} {292} (\bibinfo {year} {2011})}\BibitemShut {NoStop}%
\bibitem [{\citenamefont {Clanet}\ \emph {et~al.}(2004)\citenamefont {Clanet},
  \citenamefont {B{\'e}guin}, \citenamefont {Richard}, \citenamefont
  {Qu{\'e}r{\'e}} \emph {et~al.}}]{clanet2004maximal}%
  \BibitemOpen
  \bibfield  {author} {\bibinfo {author} {\bibfnamefont {C.}~\bibnamefont
  {Clanet}}, \bibinfo {author} {\bibfnamefont {C.}~\bibnamefont {B{\'e}guin}},
  \bibinfo {author} {\bibfnamefont {D.}~\bibnamefont {Richard}}, \bibinfo
  {author} {\bibfnamefont {D.}~\bibnamefont {Qu{\'e}r{\'e}}},  \emph {et~al.},\
  }\href@noop {} {\bibfield  {journal} {\bibinfo  {journal} {J.~Fluid Mech.}\
  }\textbf {\bibinfo {volume} {517}},\ \bibinfo {pages} {199} (\bibinfo {year}
  {2004})}\BibitemShut {NoStop}%
\bibitem [{\citenamefont {Peters}\ \emph {et~al.}(2013)\citenamefont {Peters},
  \citenamefont {Xu},\ and\ \citenamefont {Jaeger}}]{peters2013does}%
  \BibitemOpen
  \bibfield  {author} {\bibinfo {author} {\bibfnamefont {I.~R.}\ \bibnamefont
  {Peters}}, \bibinfo {author} {\bibfnamefont {Q.}~\bibnamefont {Xu}}, \ and\
  \bibinfo {author} {\bibfnamefont {H.~M.}\ \bibnamefont {Jaeger}},\ }\href
  {\doibase 10.1103/PhysRevLett.111.028301} {\bibfield  {journal} {\bibinfo
  {journal} {Phys. Rev. Lett.}\ }\textbf {\bibinfo {volume} {111}},\ \bibinfo
  {pages} {028301} (\bibinfo {year} {2013})}\BibitemShut {NoStop}%
\bibitem [{\citenamefont {de~Jong}\ \emph {et~al.}(2007)\citenamefont
  {de~Jong}, \citenamefont {Reinten}, \citenamefont {Wijshoff}, \citenamefont
  {van~den Berg}, \citenamefont {Delescen}, \citenamefont {van Dongen},
  \citenamefont {Mugele}, \citenamefont {Versluis},\ and\ \citenamefont
  {Lohse}}]{de2007marangoni}%
  \BibitemOpen
  \bibfield  {author} {\bibinfo {author} {\bibfnamefont {J.}~\bibnamefont
  {de~Jong}}, \bibinfo {author} {\bibfnamefont {H.}~\bibnamefont {Reinten}},
  \bibinfo {author} {\bibfnamefont {H.}~\bibnamefont {Wijshoff}}, \bibinfo
  {author} {\bibfnamefont {M.}~\bibnamefont {van~den Berg}}, \bibinfo {author}
  {\bibfnamefont {K.}~\bibnamefont {Delescen}}, \bibinfo {author}
  {\bibfnamefont {R.}~\bibnamefont {van Dongen}}, \bibinfo {author}
  {\bibfnamefont {F.}~\bibnamefont {Mugele}}, \bibinfo {author} {\bibfnamefont
  {M.}~\bibnamefont {Versluis}}, \ and\ \bibinfo {author} {\bibfnamefont
  {D.}~\bibnamefont {Lohse}},\ }\href@noop {} {\bibfield  {journal} {\bibinfo
  {journal} {Appl.~Phys.~Lett.}\ }\textbf {\bibinfo {volume} {91}},\ \bibinfo
  {pages} {204102} (\bibinfo {year} {2007})}\BibitemShut {NoStop}%
\bibitem [{\citenamefont {Hu}\ and\ \citenamefont
  {Larson}(2006)}]{hu2006marangoni}%
  \BibitemOpen
  \bibfield  {author} {\bibinfo {author} {\bibfnamefont {H.}~\bibnamefont
  {Hu}}\ and\ \bibinfo {author} {\bibfnamefont {R.~G.}\ \bibnamefont
  {Larson}},\ }\href@noop {} {\bibfield  {journal} {\bibinfo  {journal}
  {J.~Phys.~Chem.~B}\ }\textbf {\bibinfo {volume} {110}},\ \bibinfo {pages}
  {7090} (\bibinfo {year} {2006})}\BibitemShut {NoStop}%
\bibitem [{\citenamefont {Nikolov}\ \emph {et~al.}(2002)\citenamefont
  {Nikolov}, \citenamefont {Wasan}, \citenamefont {Chengara}, \citenamefont
  {Koczo}, \citenamefont {Policello},\ and\ \citenamefont
  {Kolossvary}}]{nikolov2002superspreading}%
  \BibitemOpen
  \bibfield  {author} {\bibinfo {author} {\bibfnamefont {A.~D.}\ \bibnamefont
  {Nikolov}}, \bibinfo {author} {\bibfnamefont {D.~T.}\ \bibnamefont {Wasan}},
  \bibinfo {author} {\bibfnamefont {A.}~\bibnamefont {Chengara}}, \bibinfo
  {author} {\bibfnamefont {K.}~\bibnamefont {Koczo}}, \bibinfo {author}
  {\bibfnamefont {G.~A.}\ \bibnamefont {Policello}}, \ and\ \bibinfo {author}
  {\bibfnamefont {I.}~\bibnamefont {Kolossvary}},\ }\href@noop {} {\bibfield
  {journal} {\bibinfo  {journal} {Adv.~Colloid~Interface~Sci.}\ }\textbf
  {\bibinfo {volume} {96}},\ \bibinfo {pages} {325} (\bibinfo {year}
  {2002})}\BibitemShut {NoStop}%
\bibitem [{\citenamefont {Nicolas}(2005)}]{nicolas2005spreading}%
  \BibitemOpen
  \bibfield  {author} {\bibinfo {author} {\bibfnamefont {M.}~\bibnamefont
  {Nicolas}},\ }\href@noop {} {\bibfield  {journal} {\bibinfo  {journal}
  {J.~Fluid~Mech.~}\ }\textbf {\bibinfo {volume} {545}},\ \bibinfo {pages}
  {271} (\bibinfo {year} {2005})}\BibitemShut {NoStop}%
\bibitem [{mov()}]{movie}%
  \BibitemOpen
  \href@noop {} {}\bibinfo {note} {See the experimental movie.}\BibitemShut
  {Stop}%
\bibitem [{\citenamefont {Qi}\ \emph {et~al.}(2011)\citenamefont {Qi},
  \citenamefont {McMurry}, \citenamefont {Norris},\ and\ \citenamefont
  {Girshick}}]{qi2011impact}%
  \BibitemOpen
  \bibfield  {author} {\bibinfo {author} {\bibfnamefont {L.}~\bibnamefont
  {Qi}}, \bibinfo {author} {\bibfnamefont {P.~H.}\ \bibnamefont {McMurry}},
  \bibinfo {author} {\bibfnamefont {D.~J.}\ \bibnamefont {Norris}}, \ and\
  \bibinfo {author} {\bibfnamefont {S.~L.}\ \bibnamefont {Girshick}},\
  }\href@noop {} {\bibfield  {journal} {\bibinfo  {journal} {Langmuir}\
  }\textbf {\bibinfo {volume} {27}},\ \bibinfo {pages} {12677} (\bibinfo {year}
  {2011})}\BibitemShut {NoStop}%
\bibitem [{\citenamefont {Fukumoto}\ and\ \citenamefont
  {Huang}(1999)}]{fukumoto1999flattening}%
  \BibitemOpen
  \bibfield  {author} {\bibinfo {author} {\bibfnamefont {M.}~\bibnamefont
  {Fukumoto}}\ and\ \bibinfo {author} {\bibfnamefont {Y.}~\bibnamefont
  {Huang}},\ }\href@noop {} {\bibfield  {journal} {\bibinfo  {journal}
  {J.~Thermal~Spray~Tech.}\ }\textbf {\bibinfo {volume} {8}},\ \bibinfo {pages}
  {427} (\bibinfo {year} {1999})}\BibitemShut {NoStop}%
\bibitem [{\citenamefont {Seerden}\ \emph {et~al.}(2004)\citenamefont
  {Seerden}, \citenamefont {Reis}, \citenamefont {Evans}, \citenamefont
  {Grant}, \citenamefont {Halloran},\ and\ \citenamefont
  {Derby}}]{seerden2004ink}%
  \BibitemOpen
  \bibfield  {author} {\bibinfo {author} {\bibfnamefont {K.~A.}\ \bibnamefont
  {Seerden}}, \bibinfo {author} {\bibfnamefont {N.}~\bibnamefont {Reis}},
  \bibinfo {author} {\bibfnamefont {J.~R.}\ \bibnamefont {Evans}}, \bibinfo
  {author} {\bibfnamefont {P.~S.}\ \bibnamefont {Grant}}, \bibinfo {author}
  {\bibfnamefont {J.~W.}\ \bibnamefont {Halloran}}, \ and\ \bibinfo {author}
  {\bibfnamefont {B.}~\bibnamefont {Derby}},\ }\href@noop {} {\bibfield
  {journal} {\bibinfo  {journal} {J.~Am.~Ceram.~Soc.}\ }\textbf {\bibinfo
  {volume} {84}},\ \bibinfo {pages} {2514} (\bibinfo {year}
  {2004})}\BibitemShut {NoStop}%
\bibitem [{\citenamefont {Derby}(2011)}]{derby2011inkjet}%
  \BibitemOpen
  \bibfield  {author} {\bibinfo {author} {\bibfnamefont {B.}~\bibnamefont
  {Derby}},\ }\href@noop {} {\bibfield  {journal} {\bibinfo  {journal}
  {J.~Eur.~Ceram.~Soc.}\ }\textbf {\bibinfo {volume} {31}},\ \bibinfo {pages}
  {2543} (\bibinfo {year} {2011})}\BibitemShut {NoStop}%
\bibitem [{\citenamefont {Miskin}\ and\ \citenamefont
  {Jaeger}(2012)}]{miskin2012droplet}%
  \BibitemOpen
  \bibfield  {author} {\bibinfo {author} {\bibfnamefont {Z.}~\bibnamefont
  {Miskin}}\ and\ \bibinfo {author} {\bibfnamefont {H.~M.}\ \bibnamefont
  {Jaeger}},\ }\href@noop {} {\bibfield  {journal} {\bibinfo  {journal}
  {Proc.~Natl.~Acad.~Sci.}\ }\textbf {\bibinfo {volume} {109}},\ \bibinfo
  {pages} {4389} (\bibinfo {year} {2012})}\BibitemShut {NoStop}%
\bibitem [{\citenamefont {Bertrand}\ \emph {et~al.}(2012)\citenamefont
  {Bertrand}, \citenamefont {Bonnoit}, \citenamefont {Cl{\'e}ment},\ and\
  \citenamefont {Lindner}}]{bertrand2012dynamics}%
  \BibitemOpen
  \bibfield  {author} {\bibinfo {author} {\bibfnamefont {T.}~\bibnamefont
  {Bertrand}}, \bibinfo {author} {\bibfnamefont {C.}~\bibnamefont {Bonnoit}},
  \bibinfo {author} {\bibfnamefont {E.}~\bibnamefont {Cl{\'e}ment}}, \ and\
  \bibinfo {author} {\bibfnamefont {A.}~\bibnamefont {Lindner}},\ }\href@noop
  {} {\bibfield  {journal} {\bibinfo  {journal} {Granul. Matter}\ }\textbf
  {\bibinfo {volume} {14}},\ \bibinfo {pages} {169} (\bibinfo {year}
  {2012})}\BibitemShut {NoStop}%
\bibitem [{\citenamefont {Yarin}\ and\ \citenamefont
  {Weiss}(1995)}]{yarin1995impact}%
  \BibitemOpen
  \bibfield  {author} {\bibinfo {author} {\bibfnamefont {A.}~\bibnamefont
  {Yarin}}\ and\ \bibinfo {author} {\bibfnamefont {D.}~\bibnamefont {Weiss}},\
  }\href@noop {} {\bibfield  {journal} {\bibinfo  {journal} {J. Fluid Mech.}\
  }\textbf {\bibinfo {volume} {283}},\ \bibinfo {pages} {141} (\bibinfo {year}
  {1995})}\BibitemShut {NoStop}%
\bibitem [{wet()}]{wettingComment}%
  \BibitemOpen
  \href@noop {} {}\bibinfo {note} {The value of $\alpha$ is insensitive to the
  contact line dynamics. Impacts using plugs suspended in silicone oil show the
  same expansion dynamics as those suspended in water even though oil wets the
  dry glass substrate but water does not. This insensitivity arises because the
  region of the liquid surface affected by the substrate wetting properties
  corresponds to a very small portion of the full area when the liquid motion
  Weber number are large and $We_p \gg 1$.}\BibitemShut {Stop}%
\bibitem [{sup()}]{supplementalMaterials}%
  \BibitemOpen
  \href@noop {} {}\bibinfo {note} {See the details in the Supplemental
  Material.}\BibitemShut {Stop}%
\bibitem [{Note1()}]{Note1}%
  \BibitemOpen
  \bibinfo {note} {Varying $\alpha $ by 10\% still allows a fit within the
  error bar range of experiments.}\BibitemShut {Stop}%
\bibitem [{\citenamefont {Vella}\ and\ \citenamefont
  {Mahadevan}(2005)}]{vella2005cheerios}%
  \BibitemOpen
  \bibfield  {author} {\bibinfo {author} {\bibfnamefont {D.}~\bibnamefont
  {Vella}}\ and\ \bibinfo {author} {\bibfnamefont {L.}~\bibnamefont
  {Mahadevan}},\ }\href@noop {} {\bibfield  {journal} {\bibinfo  {journal}
  {Am.~J.~Phys.}\ }\textbf {\bibinfo {volume} {73}},\ \bibinfo {pages} {817}
  (\bibinfo {year} {2005})}\BibitemShut {NoStop}%
\bibitem [{\citenamefont {Herminghaus}(2005)}]{herminghaus2005dynamics}%
  \BibitemOpen
  \bibfield  {author} {\bibinfo {author} {\bibfnamefont {S.}~\bibnamefont
  {Herminghaus}},\ }\href@noop {} {\bibfield  {journal} {\bibinfo  {journal}
  {Adv. Phys.}\ }\textbf {\bibinfo {volume} {54}},\ \bibinfo {pages} {221}
  (\bibinfo {year} {2005})}\BibitemShut {NoStop}%
\bibitem [{\citenamefont {Ward}\ \emph {et~al.}(2009)\citenamefont {Ward},
  \citenamefont {Wey}, \citenamefont {Glidden}, \citenamefont {Hosoi},\ and\
  \citenamefont {Bertozzi}}]{ward2009experimental}%
  \BibitemOpen
  \bibfield  {author} {\bibinfo {author} {\bibfnamefont {T.}~\bibnamefont
  {Ward}}, \bibinfo {author} {\bibfnamefont {C.}~\bibnamefont {Wey}}, \bibinfo
  {author} {\bibfnamefont {R.}~\bibnamefont {Glidden}}, \bibinfo {author}
  {\bibfnamefont {A.~E.}\ \bibnamefont {Hosoi}}, \ and\ \bibinfo {author}
  {\bibfnamefont {A.}~\bibnamefont {Bertozzi}},\ }\href@noop {} {\bibfield
  {journal} {\bibinfo  {journal} {Phys.~Fluids}\ }\textbf {\bibinfo {volume}
  {21}},\ \bibinfo {pages} {083305} (\bibinfo {year} {2009})}\BibitemShut
  {NoStop}%
\bibitem [{\citenamefont {Timberlake}\ and\ \citenamefont
  {Morris}(2005)}]{timberlake2005particle}%
  \BibitemOpen
  \bibfield  {author} {\bibinfo {author} {\bibfnamefont {B.~D.}\ \bibnamefont
  {Timberlake}}\ and\ \bibinfo {author} {\bibfnamefont {J.~F.}\ \bibnamefont
  {Morris}},\ }\href@noop {} {\bibfield  {journal} {\bibinfo  {journal}
  {J.~Fluid~Mech.}\ }\textbf {\bibinfo {volume} {538}},\ \bibinfo {pages} {309}
  (\bibinfo {year} {2005})}\BibitemShut {NoStop}%
\bibitem [{\citenamefont {Colosqui}\ \emph {et~al.}(2013)\citenamefont
  {Colosqui}, \citenamefont {Morris},\ and\ \citenamefont
  {Stone}}]{colosqui2013hydrodynamically}%
  \BibitemOpen
  \bibfield  {author} {\bibinfo {author} {\bibfnamefont {C.~E.}\ \bibnamefont
  {Colosqui}}, \bibinfo {author} {\bibfnamefont {J.~F.}\ \bibnamefont
  {Morris}}, \ and\ \bibinfo {author} {\bibfnamefont {H.~A.}\ \bibnamefont
  {Stone}},\ }\href@noop {} {\bibfield  {journal} {\bibinfo  {journal}
  {Phys.~Rev.~Lett.}\ }\textbf {\bibinfo {volume} {110}},\ \bibinfo {pages}
  {188302} (\bibinfo {year} {2013})}\BibitemShut {NoStop}%
\bibitem [{\citenamefont {Guazzelli}\ and\ \citenamefont
  {Hinch}(2011)}]{guazzelli2011fluctuations}%
  \BibitemOpen
  \bibfield  {author} {\bibinfo {author} {\bibfnamefont {{\'E}.}~\bibnamefont
  {Guazzelli}}\ and\ \bibinfo {author} {\bibfnamefont {J.}~\bibnamefont
  {Hinch}},\ }\href@noop {} {\bibfield  {journal} {\bibinfo  {journal}
  {Annu.~Rev.~Fluid~Mech.}\ }\textbf {\bibinfo {volume} {43}},\ \bibinfo
  {pages} {97} (\bibinfo {year} {2011})}\BibitemShut {NoStop}%
\bibitem [{\citenamefont {Stickel}\ and\ \citenamefont
  {Powell}(2005)}]{stickelPowellreview05}%
  \BibitemOpen
  \bibfield  {author} {\bibinfo {author} {\bibfnamefont {J.~J.}\ \bibnamefont
  {Stickel}}\ and\ \bibinfo {author} {\bibfnamefont {R.~L.}\ \bibnamefont
  {Powell}},\ }\href@noop {} {\bibfield  {journal} {\bibinfo  {journal} {Annu.
  Rev. Fluid Mech.}\ }\textbf {\bibinfo {volume} {37}},\ \bibinfo {pages} {129}
  (\bibinfo {year} {2005})}\BibitemShut {NoStop}%
\bibitem [{\citenamefont {Lerner}\ \emph {et~al.}(2012)\citenamefont {Lerner},
  \citenamefont {D{\"u}ring},\ and\ \citenamefont {Wyart}}]{mwyartpnas12}%
  \BibitemOpen
  \bibfield  {author} {\bibinfo {author} {\bibfnamefont {E.}~\bibnamefont
  {Lerner}}, \bibinfo {author} {\bibfnamefont {G.}~\bibnamefont {D{\"u}ring}},
  \ and\ \bibinfo {author} {\bibfnamefont {M.}~\bibnamefont {Wyart}},\
  }\href@noop {} {\bibfield  {journal} {\bibinfo  {journal}
  {Proc.~Natl.~Acad.~Sci.}\ }\textbf {\bibinfo {volume} {109}},\ \bibinfo
  {pages} {4798} (\bibinfo {year} {2012})}\BibitemShut {NoStop}%
\bibitem [{\citenamefont {Brader}\ \emph {et~al.}(2008)\citenamefont {Brader},
  \citenamefont {Cates},\ and\ \citenamefont {Fuchs}}]{catesPRL09}%
  \BibitemOpen
  \bibfield  {author} {\bibinfo {author} {\bibfnamefont {J.}~\bibnamefont
  {Brader}}, \bibinfo {author} {\bibfnamefont {M.}~\bibnamefont {Cates}}, \
  and\ \bibinfo {author} {\bibfnamefont {M.}~\bibnamefont {Fuchs}},\
  }\href@noop {} {\bibfield  {journal} {\bibinfo  {journal} {Phys.~Rev.~Lett.}\
  }\textbf {\bibinfo {volume} {101}},\ \bibinfo {pages} {138301} (\bibinfo
  {year} {2008})}\BibitemShut {NoStop}%
\bibitem [{\citenamefont {Bonnoit}\ \emph {et~al.}(2010)\citenamefont
  {Bonnoit}, \citenamefont {Lanuza}, \citenamefont {Lindner},\ and\
  \citenamefont {Clement}}]{clementMesoscale}%
  \BibitemOpen
  \bibfield  {author} {\bibinfo {author} {\bibfnamefont {C.}~\bibnamefont
  {Bonnoit}}, \bibinfo {author} {\bibfnamefont {J.}~\bibnamefont {Lanuza}},
  \bibinfo {author} {\bibfnamefont {A.}~\bibnamefont {Lindner}}, \ and\
  \bibinfo {author} {\bibfnamefont {E.}~\bibnamefont {Clement}},\ }\href@noop
  {} {\bibfield  {journal} {\bibinfo  {journal} {Phys.~Rev.~Lett.}\ }\textbf
  {\bibinfo {volume} {105}},\ \bibinfo {pages} {108302} (\bibinfo {year}
  {2010})}\BibitemShut {NoStop}%
\bibitem [{\citenamefont {Mewis}\ and\ \citenamefont
  {Wagner}(2009)}]{wagnerRheologyreview}%
  \BibitemOpen
  \bibfield  {author} {\bibinfo {author} {\bibfnamefont {J.}~\bibnamefont
  {Mewis}}\ and\ \bibinfo {author} {\bibfnamefont {N.~J.}\ \bibnamefont
  {Wagner}},\ }\href@noop {} {\bibfield  {journal} {\bibinfo  {journal}
  {J.~Non-Newton.~Fluid~Mech.}\ }\textbf {\bibinfo {volume} {157}},\ \bibinfo
  {pages} {147} (\bibinfo {year} {2009})}\BibitemShut {NoStop}%
\bibitem [{\citenamefont {Boyer}\ \emph {et~al.}(2011)\citenamefont {Boyer},
  \citenamefont {Guazzelli},\ and\ \citenamefont {Pouliquen}}]{pouliquenPRL}%
  \BibitemOpen
  \bibfield  {author} {\bibinfo {author} {\bibfnamefont {F.}~\bibnamefont
  {Boyer}}, \bibinfo {author} {\bibfnamefont {{\'E}.}~\bibnamefont
  {Guazzelli}}, \ and\ \bibinfo {author} {\bibfnamefont {O.}~\bibnamefont
  {Pouliquen}},\ }\href@noop {} {\bibfield  {journal} {\bibinfo  {journal}
  {Phys.~Rev.~Lett.}\ }\textbf {\bibinfo {volume} {107}},\ \bibinfo {pages}
  {188301} (\bibinfo {year} {2011})}\BibitemShut {NoStop}%
\bibitem [{\citenamefont {Koos}\ and\ \citenamefont
  {Willenbacher}(2011)}]{capillarySuspensionSci12}%
  \BibitemOpen
  \bibfield  {author} {\bibinfo {author} {\bibfnamefont {E.}~\bibnamefont
  {Koos}}\ and\ \bibinfo {author} {\bibfnamefont {N.}~\bibnamefont
  {Willenbacher}},\ }\href@noop {} {\bibfield  {journal} {\bibinfo  {journal}
  {Science}\ }\textbf {\bibinfo {volume} {331}},\ \bibinfo {pages} {897}
  (\bibinfo {year} {2011})}\BibitemShut {NoStop}%
\bibitem [{\citenamefont {Brown}\ and\ \citenamefont
  {Jaeger}(2012)}]{ebrownJRheo12}%
  \BibitemOpen
  \bibfield  {author} {\bibinfo {author} {\bibfnamefont {E.}~\bibnamefont
  {Brown}}\ and\ \bibinfo {author} {\bibfnamefont {H.~M.}\ \bibnamefont
  {Jaeger}},\ }\href@noop {} {\bibfield  {journal} {\bibinfo  {journal}
  {J.~Rheol.}\ }\textbf {\bibinfo {volume} {56}},\ \bibinfo {pages} {875}
  (\bibinfo {year} {2012})}\BibitemShut {NoStop}%
\end{thebibliography}

\end{document}